\documentclass[prd,twocolumn,floats,superscriptaddress,eqsecnum,floatfix,
showpacs,longbibliography]{revtex4-1}

\usepackage{amssymb,amsmath}
\usepackage{epsfig}

\pdfoutput=1
\usepackage{amssymb,amsmath}
\usepackage{epsfig}
\usepackage{color}
\usepackage{datetime}
\usepackage[utf8]{inputenc}

\newcommand{\eps}{\varepsilon}

\newcommand{\pa}{\partial}

\newcommand{\be}{\begin{equation}}
\newcommand{\BE}[1]{\begin{equation} \n{#1}}
\newcommand{\ee}{\end{equation}}
\newcommand{\ba}{\begin{eqnarray}}
\newcommand{\ea}{\end{eqnarray}}
\newcommand{\beg}{\begin{gather*}}
\newcommand{\eng}{\end{gather*}}

\newcommand{\hh}{,\hspace{0.5cm}}
\newcommand{\hhh}{,\hspace{0.2cm}}

\newcommand{\eq}[1]{(\ref{#1})}

\newcommand{\n}[1]{\label{#1}}

\newcommand{\hor}{\stackrel{ {\mbox{\tiny H}}}{=} }

\def\XXint#1#2#3{{\setbox0=\hbox{$#1{#2#3}{\int}$ }
\vcenter{\hbox{$#2#3$ }}\kern-.6\wd0}}

\begin{document}

\title{Notes on non-singular models of black holes}

\author{Valeri P. Frolov}
\email{vfrolov@ualberta.ca}
\affiliation{Theoretical Physics Institute, Department of Physics\\
University of Alberta, Edmonton, Alberta, Canada T6G 2E1}

%\today, \currenttime

\begin{abstract}
We discuss static spherically symmetric metrics which represent non-singular black holes in four- and higher-dimensional spacetime. We impose a set of restrictions, such as a regularity of the metric at the center $r=0$ and Schwarzschild asymptotic behavior at large $r$. We assume that the metric besides mass $M$ contains an additional parameter $\ell$, which determines the scale where modification of the solution of the Einstein equations becomes significant. We require that the modified metric obeys the limiting curvature condition, that is its curvature is uniformly restricted by the value $\sim \ell^{-2}$. We also make a "more technical" assumption that the metric coefficients are rational functions of $r$. In particular, the invariant $(\nabla r)^2$ has the form $P_n(r)/\tilde{P}_n(r)$, where $P_n$ and $\tilde{P}_n$ are polynomials of the order of $n$. We discuss first the case of four dimensions. We show that  when $n\le 2$ such a metric cannot describe a non-singular black hole. For $n=3$ we find a suitable metric, which besides $M$ and $\ell$ contains a dimensionless numerical parameter. When this parameter vanishes the obtained metric coincides with Hayward's one. The characteristic property of such spacetimes is $-\xi^2=(\nabla r)^2$, where $\xi^2$ is a time-like at infinity Killing vector. We describe a possible generalization of a non-singular black-hole metric to the case when this equality is violated. We also obtain a metric for a charged non-singular black hole obeying similar restrictions as the neutral one, and construct higher dimensional models of neutral and charged black holes.
\end{abstract}


\maketitle

%%%%%%%%%%%%%%%%%%%%%%%%%%%%%%%%%%%%%%%%%%%%%%%%%%%%%%%%%%%%

%%
\section{Introduction}

The General Relativity is ultraviolet-incomplete, both in the classical and quantum domains. A well-known problem of the classical Einstein’s theory of gravity is inevitable existence of singularities. For example, solutions of the Einstein equations, describing stationary isolated black holes, such as Schwarzschild, Reisner-Nordstr\"{o}m and Kerr metric, have curvature singularity in their interior. It is generally believed that the General Relativity should be modified in the regions where the spacetime curvature becomes high. Such a modification is also required if one searches for a theory which is ultraviolet (UV) complete. There exist several different proposals how such a modification can be achieved. For example, quite long time ago it has been demonstrated that addition of the higher order in curvature terms, as well as the terms containing higher derivatives, can improve the UV properties of the Einstein gravity \cite{Stelle:1977ry,Stelle:1976gc,Biswas:2005qr,Barnaby:2007ve}. However such theories usually have non-physical degrees of freedom (ghosts). Recently a new version of UV complete modification of the General Relativity was proposed which is free from this problem \cite{Tomboulis:1997gg, Biswas:2011ar, Biswas:2013cha}. It was named a ghost-free gravity \cite{Tomboulis:1997gg, Biswas:2011ar, Modesto:2011kw, Modesto:2012ys, Biswas:2013cha, Biswas:2013kla,
Modesto:2014lga, Talaganis:2014ida,Tomboulis:2015gfa,Tomboulis:2015esa}. Such a theory contains an infinite number of derivatives and, in fact, is non-local \cite{Tomboulis:2015gfa, Biswas:2013kla}.
Similar theory   appears naturally also in the context of noncommutative geometry deformation of the
Einstein gravity \cite{Nicolini:2005vd,Spallucci:2006zj} (see a review \cite{Nicolini:2005zi} and references therein). The application of the ghost free theory of gravity to the
problem of singularities in cosmology and black holes can be found in
\cite{Biswas:2010zk, Modesto:2010uh, Hossenfelder:2009fc, Calcagni:2013vra,
Zhang:2014bea, Conroy:2015wfa,Li:2015bqa,Bambi:2016uda}.

In the absence of the adopted theory it is instructive to discuss what kind of modifications one could expect when gravity is UV complete. Such an analysis can be fruitful only if some "natural" assumptions concerning the properties of such "full" theory are imposed. In the present paper we present some results on so-called non-singular models of black holes.

One of the main assumptions is that there exists a critical energy $\mu$ and the corresponding length-scale parameter $\ell=\mu^{-1}$. The metric should be modified when the spacetime curvature ${\cal R}$ become comparable with $\ell^{-2}$. At the same time we assume that one can use the classical metric $g_{\mu\nu}$, which is a solution of the effective action of the modified gravity. In other words, the length scale $\lambda$, where quantum gravity effects become important, is much smaller than $\ell$. We are looking for black-hole metrics which do not have curvature singularities. The first model of a non-singular black hole was proposed by Bardeen \cite{Bardeen:1968}, who considered a collapse of a charged matter with a charged matter core inside the black hole instead of its singularity. Different models of neutral, charged and rotating non-singular black holes were proposed and discussed later \cite{Dymnikova:1992ux,Borde,AyonBeato:1998ub,Lemos:2011dq,Uchikata:2012zs,Flachi:2012nv,DeLorenzo:2014pta,
Balart:2014cga,Ghosh:2014hea,DeLorenzo:2015taa,Rovelli,Kawai,Lemos:2016ulj}. A general review of different models of non-singular black holes and additional references on this subject can be found in \cite{Ansoldi:2008jw}.

In a general case a regular solution besides some critical scale parameter $\ell$, which is a parameter of the corresponding UV complete theory, contains also such parameters as mass and charge, which specify a concrete solution. The regularity of the solution means that for a fixed value of these parameters the curvature of the spacetime is finite. However, in a general case there is no guarantee that the maximal value of the curvature would not infinitely grow when, say, the mass (and/or charge) becomes infinitely large. It is reasonable to assume that for a viable fundamental theory the absolute maximal value of the curvature is restricted by some fundamental value. In other words, the curvature invariants are uniformly restricted by some universal value
\BE{0.1}
|{\cal R}|\le {c\over \ell^2}\, ,
\ee
where $c$ is a dimensionless constant which depends only on the type of the curvature invariant.
This assumption, called the limiting curvature conjecture, was proposed in  \cite{Markov:1982,Markov:1984ii,Polchinski:1989ae}.
In the present paper we assume that the limiting curvature condition is satisfied.
To make this condition more concrete, we shall check restriction \eq{0.1} for the Ricci scalar, as well for the square roots of the quadratic in the curvature invariants. We shall demonstrate that the limiting curvature condition imposes a significant restriction on regular metrics. Many of the proposed earlier non-singular black-hole models violate this property.

In the present paper we consider  spherically symmetric non-singular black holes. This simplifying assumption allows one to make many results more concrete. For example, simple analysis of the spherically symmetric metrics shows that if such a spacetime has an apparent horizon it cannot cross the center $r=0$ without creation of the curvature singularity  (see next Section). This means that besides an outer horizon, close to the classical gravitational radius,  such a regular metric has also an inner horizon, close to $r=0$.  These two horizons either never meet in the future, or the apparent horizon is closed \cite{Frolov:1981mz}.
New (baby) universe creation inside a black hole, discussed in \cite{Frolov:1989pf,Frolov:1988vj}, is an example of the former case. (See also \cite{Morgan:1990yy, Barrabes:1995nk,
Easson:2001qf,Dymnikova:2004qg,Lukash:2013ts,Dymnikova:2016nlb}). A model, describing a complete quantum evaporation of a non-singular black hole, described in \cite{Frolov:1981mz}, is an example of the latter case. Such models were later intensively discussed in the literature \cite{Roman:1983zza,Hayward:2005gi,Hossenfelder:2009fc,Bambi:2013gva,Bambi:2013caa,Hawking:2014tga,
Frolov:2014jva,Bardeen:2014uaa,Barrau:2015uca}. Non-singular black holes models in the dilaton 2D gravity were discussed in \cite{Trodden:1993dm,Easson:2002tg,Taves:2014laa, Kunstatter:2015raa,Kunstatter:2015vxa}.

A special interest to a  non-singular model of a completely evaporating black-hole is connected with a long-standing problem of the information loss in black holes. In case when the apparent horizon is closed, the event horizon is absent, and all the information, accumulated inside such a "black hole", can return to the spacetime, visible by an external observer, after the evaporation is completed (see e.g. \cite{Frolov:2014jva} and references there in).

A convenient for the analysis model of a non-singular black hole was proposed by Hayward \cite{Hayward:2005gi} (see also \cite{Frolov:2014jva}). It describes an isolated four-dimensional spherically symmetric regular spacetime and, besides the fundamental length $\ell$, contains only one parameter, mass $M$. At large distance it reproduces the Schwarzschild metric, while at the origin it is regular and has de Sitter form. For $M\ge 3\sqrt{3} \ell/4$ the metric has two branches (outer and inner) of the apparent horizon. The outer horizon is located near $2M$, while the inner one is close to $\ell$. The property which makes this metric simple for the analysis is its scaling behavior. Namely, there exist a scaling transformation of the coordinates, metric and its parameters, which preserves the form of the metric. We discuss this property in the Section~II.

In the present paper we propose and discuss useful generalizations of the Hayward metric.
A first important generalization is a wide class of metrics, with non-trivial red-shift factor.
We also present a higher dimension generalization of a  non-singular  metric,  as well as  metrics for charged non-singular black hole spacetimes.

\section{Metric of a non-singular black hole}

\subsection{A non-singular black-hole model}

A general   static metric  in a four-dimensional spacetime  can be written in the form
\be\n{a.1}
dS^2=-F A^2 dV^2+2 A dV dr+r^2 d\omega^2\, ,
\ee
where  $F=F(r)$ and $A=A(r)$ are two arbitrary functions. This metric has the Killing vector $\xi^{\mu}\pa_{\mu}=\pa_V$.
 It is easy to see that
\BE{a.1b}
F=(\nabla r)^2\hh F A^2 =-\xi^2\, .
\ee
In a spacetime with a horizon, $F(r)$ vanishes at the position $r_0$ of the apparent horizon. For a regular static metric such a horizon is at the same time the Killing horizon, so that $A(r_0)$ is finite there.

If the metric has a horizon where $F(r_0)=0$ then
\be\n{a.1a}
\xi^{\nu}\xi^{\mu}_{\ ;\nu}\hor \kappa \xi^{\mu}\hh \kappa={1\over 2}(A F')|_{r=r_0}\, .
\ee
By definition, $\kappa$ is the surface gravity. The value of $\kappa$ depends on the choice of the normalization of the Killing vector.
In an asymptotically flat spacetime one usually puts $\xi^2|_{r=\infty}=-1$.  A conditions that there is no solid angle deficit implies $F|_{\infty}=1$. Hence one also has $A|_{\infty}=1$.

Let $R$ be the Ricci scalar, $S_{\mu\nu}=R_{\mu\nu}-{1\over 4}g_{\mu\nu}R $, and $C_{\mu\nu\alpha\beta} $ be the Weyl tensor. Let us define the following quadratic in the curvature invariants
\be\n{a.4}
{\cal S}^2=S_{\mu\nu} S^{\mu\nu}\hh
{\cal C}^2=C_{\mu\nu\alpha\beta} C^{\mu\nu\alpha\beta}\, .
\ee
Then one has
\ba\n{a.5}
R&=& F''+{4\over r}F'-2{F-1\over r^2}\nonumber\\
&+&{1\over A}\left(2FA''+3 F' A' +{4\over r} F A'\right) \, ,\\
{\cal C}&=&{1\over \sqrt{3}}\left[ F'' -{2\over r}F'+2{F-1\over r^2}\right.\nonumber\\
&+&\left. {1\over A}\left(2F A''+3 F' A' -{2\over r} F A'\right) \right]\, .
\ea
An expression for ${\cal S}$ is similar but quite long and we do not present it here.

We assume that the metric \eq{a.1} is finite at the origin $r=0$, so that
\ba
F&=&F_0+F_1 r+F_2 r^2+O(r^3)\, ,\nonumber\\
A&=&A_0+A_1 r+A_2 r^2+O(r^3)\, .\n{a.3}
\ea
In a general case the curvature invariants $R$ and ${\cal C}$ are singular at the origin and have divergences $\sim r^{-2}$ and $\sim r^{-1}$. Conditions that these divergences are absent and the metric is regular are
\BE{a.3a}
F_0=1\hh F_1=A_1=0\, .
\ee
By substituting \eq{a.3} in the expression for ${\cal S}^2$ and using relations \eq{a.3a} one can check that the condition of regularity of ${\cal S}^2$ at $r=0$ are identically satisfied. One also has
\ba\n{a.3b}
R&=&-12\left( F_2+{A_2\over A_0}\right)+O(r^2)\, ,\\
{\cal S}&=&2\sqrt{3} {A_2\over A_0}+O(r^2)\hhh
{\cal C}=O(r^2)\, .
\ea

Let us notice that $A_0$ is an arbitrary constant. Its meaning is connected with a time delay between infinity and $r=0$. For a fixed value of $r$ at far distance one has $\Delta \tau_{\infty}\equiv \Delta t=\Delta V$, where $\tau_{\infty}$ is the proper time measured by the clocks at the infinity. For the same interval $\Delta V$ the proper time measured at the center $r=0$ is
$\Delta \tau_0=A_0 \Delta V$. Hence
\BE{a.3c}
\Delta \tau_0=A_0 \Delta \tau_{\infty}\, .
\ee
For $A_0<1$ ($A_0>1$) the time at the center "goes" slower (faster) than  at the infinity. For a monochromatic wave $\varphi$ propagating in a static spacetime one can write $\varphi\sim \exp{(i \Phi)}$, where $\Phi=\omega V$. If $\omega_{\infty}=d\Phi/d \tau_{\infty}$ and $\omega_0=d\Phi/d \tau_0$, then one has
\BE{a.3d}
\omega_0=A_0^{-1} \omega_{\infty}\, .
\ee
For $A_0>1$ ($A_0<1$) the frequency of a signal, registered at the center $r=0$, is red-shifted (blue-shifted) with respect to the frequency of the signal emitted at the infinity. In what follows we refer to $A(r)$ as a red-shift function.

We are interested in a metric which describes a black hole. For this reason we assume that the function $F(r)$ vanishes at some value $r=r_+$, where the event horizon is located. In order for the metric to be regular at $r=0$ it must have at least one more zero at $r=r_->0$. For simplicity we assume that the function $F(r)$ has exactly two zeros at $r_+>r_->0$. Our final assumption is that the curvature invariants $|R|$,  $|{\cal S}|$ and $|{\cal C}|$ are uniformly restricted by some values proportional to ${\ell}^{-2}$. We call this parameter $\ell$ the fundamental length. The latter requirement means that our metric satisfies the Markov's limiting curvature conjecture \cite{Markov:1982, Markov:1984ii, Polchinski:1989ae}.

To fix the scale of the parameter $\ell$ one can put $|F_2|=\ell^{-2}$, so that the metric function $F$ at the origin $r=0$ has the form
\BE{a.7}
F=1+\eps r^2/\ell^2+O(r^4)\hh \eps=\pm 1\, .
\ee

We assume that the spacetime is asymptotically flat, so that
\be\n{a.2}
F=1-{r_g\over r}+O(r^{-2})\hh r_g=2M\, .
\ee

We call a spacetime  \eq{a.1} satisfying the above described properties (including the limiting curvature condition) a non-singular black hole. Certainly one cannot require that this metric is a solution of the Einstein equations. One should assume that the Einstein equations should be modified in the UV domain. The curvature of the Schwarzschild spacetime, $\sim r_g/r^3$, reaches the critical value $\ell^{-2}$ at the radius $r_{\ell}=(r_g \ell^2)^{1/3}$. At this  radius the modified solution becomes essentially different from the Einstein's solution.

\subsection{Uncharged non-singular black-hole metric}

\subsubsection{Scaling property}

In the absence of a "final" UV complete theory of gravity there is a wide ambiguity in the choice of metric functions $F$ and $A$ for the metric describing a modified black-hole. This ambiguity is reduced by adopting constraints described in the previous section, but it is still quite wide. We impose additional "natural" restrictions.

Let us consider the Schwarzschild metric which is a vacuum spherically symmetric solution of the Einstein equations. For this metric
\BE{a.7a}
F=1-{r_g\over r}={r-r_g\over r}\hh A=1\, .
\ee
The form of the metric is fixed by the Einstein equations and it contains one parameter, $r_g=2M$, which is the integration constant. Moreover, $F$ has the form of the rational function, which is the ratio of two first order in $r$ polynomials. For a special value of the parameter $r_g=0$ the spacetime is flat. An additional property of the Schwarzschild metric is its scale invariance. Namely, its form does not change under the following scale transformations
\BE{a.7hh}
r\to \beta r\, ,\ r_g\to \beta r_g\, ,\  v\to \beta v\, , \ ds^2\to \beta^2 ds^2\, .
\ee
This symmetry property allows one to write
\BE{a.7gg}
ds^2 =r_g^2 (ds^2)|_{r_g=1}\, .
\ee
In other words, by using the dimensional quantity $r_g$ as a general scale parameter, one reduces the original metric with one parameter (mass), to the metric $(ds^2)|_{r_g=1}$, which does not contain any arbitrary parameters at all.

\subsubsection{Case $n\le 2$}

Let us consider a generalization of this metric obeying the condition $A=1$. We assume  that $F$ is a rational function of $r$
\BE{a.7b}
F(r)={P_n(r)\over \tilde{P}_n(r)}\, ,
\ee
where $P_n$ and $\tilde{P}_n$ are polynomials of the order $n>1$. For example, one may try the following form of $F$
\BE{a.7c}
F=1-{r_g r\over r^2+\ell^2}\, .
\ee
Here the fundamental scale $\ell$ plays the role of the regularizer.  At far distance the metric correctly reproduces the Schwarzschild solution and deflection from it is of the order of $\ell^2$. At the origin the metric is finite.
However, it is not regular. Moreover, its curvature invariants have the form
\BE{a.7d}
{\cal R}={1\over \ell^2} {r_g\over \ell} f(\rho)\, ,
\ee
where $\rho=r/\ell$. For any fixed $\rho$ the corresponding curvature invariant can be made arbitrary large by simply increasing the mass parameter $r_g$. This means that the metric \eq{a.1} with \eq{a.7c} for a black hole does not satisfy the limiting curvature condition \footnote{Metrics similar \eq{a.7c} were discussed in  \cite{AyonBeato:1998ub} and related papers as a model of a regular black hole. However it looks like that all this kind of such metrics do not satisfy the limiting curvature condition.}.

Can we reach desired properties of the metric when $n=2$? One can write \eq{a.7b} for this case as follows
\BE{a.7e}
F={r^2+a_1 r+a_0\over r^2+b_1 r+b_0}\, .
\ee
Condition $F_0=1$ implies $b_0=a_0$.  It is also easy to check that the condition $F_1=0$ requires that $b_1=a_1$. But in this case $F$ is identically 1 and the spacetime is flat. To summarize, metrics \eq{a.7b} with $n\le 2$ cannot be used as a consistent model of a non-singular black hole.

\subsubsection{Case $n=3$}

Let us analyse metrics \eq{a.7b} with $n=3$
\BE{a.7h}
F={r^3+a_2 r^2+a_1 r+a_0\over r^3+b_2 r^2+b_1 r+b_0}\, .
\ee
Regularity of the spacetime at the origin $r=0$ implies
\BE{a.7g}
b_0=a_0\hh b_1=a_1\, ,
\ee
so that one has
\BE{a.7j}
F=1-{(b_2-a_2) r^2\over r^3+b_2 r^2+b_1 r+a_0}\, .
\ee
In order to have proper Schwarzschild asymptotic form one must put $b_2-a_2=r_g$. To satisfy the condition \eq{a.7} one must choose $a_0=r_g \ell^2$. Hence the metric function $F$ takes the form
\BE{a.7k}
F=1-{r_g r^2\over r^3+b_2 r^2+b_1 r+r_g \ell^2}\, .
\ee
This function, besides the fundamental length $\ell$ and the gravitational radius $r_g=2M$, contains two arbitrary parameter $b_1$ and $b_2$ with the dimensionality $[length]^2$ and $[length]$, respectively.We assume that these parameters have the form of the product of non-negative integer powers of $r_g$ and $\ell$. Then $b_2\sim \ell$ and $b_1\sim r_g \ell$ or $b_1\sim \ell^2$. The cases when $b_2\sim r_g$ and $b_1\sim r_g^2$ are excluded by the condition that in the limit $\ell\to 0$ the metric must coincides with the Schwarzschild one.

We write the metric function \eq{a.7k} as
\BE{a.7m}
F=1-{r_g r^2\over r^3+c_2 \ell r^2+(c\ell^2+c_1 \ell r_g) r+r_g \ell^2}\, .\, .
\ee
One has the following series expansion
\BE{a.7n}
F=1-{r_g\over r}+{c_2 \ell r_g\over r^2}+{\ell r_g[ (c-c_2)\ell +c_1 r_g]\over r^{3}}+O(r^{-4})\, .
\ee

In the quantum gravity the fundamental length is $\ell=\sqrt{\hbar c/G}$. Loop expansions contain integer powers of $\hbar$, or, what is equivalent, integer powers of $\ell^2$. In such case the terms linear in $\ell$ in \eq{a.7n} should vanish and one has
\BE{a.7p}
F=1-{r_g r^2\over r^3+c \ell^2 r+r_g \ell^2}\, ,
\ee
where $c$ is dimensionless numerical parameter.
Let us denote
\BE{a.7l}
\beta=(\ell/r_g)^{1/3}\hh q=(2M\ell^2)^{1/3}\hh r=q y\, .
\ee
The curvature invariants for  metric \eq{a.7p} are
\ba
R&=& {2\over \ell^2}{-y^4 z+3 y^2 z^2-3 y^3+8 y z+6\over (y^3+y z+1)^3}\, ,\nonumber\\
{\cal C}&=&{1\over \sqrt{3}\ell^2}{2 y (-3 y^5+y^3 z+6 y^2+z)\over (y^3+y z+1)^3}\, ,\n{a.7q}\\
{\cal S}&=&{1\over \ell^2}{ y (5 y^3 z+y z^2+9 y^2+2 z)\over(y^3+y z+1)^3}\, .\nonumber
\ea
Here $z=c\beta^2$. We assume that $z\ge 0$, so that the denominators in \eq{a.7q} are strictly positive.

\begin{figure}[tbp]
\centering
\includegraphics[width=8cm]{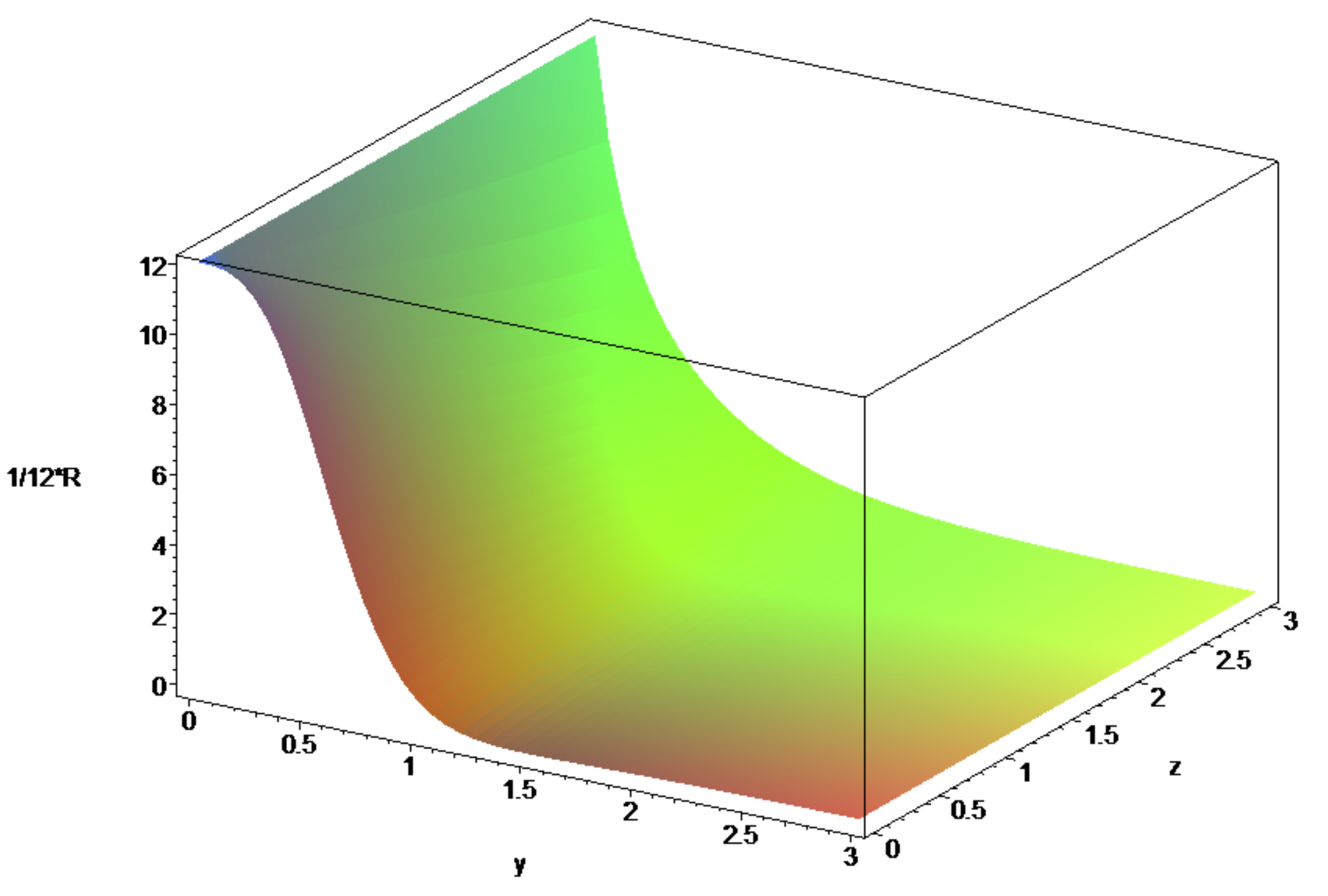}\hfill
  \caption{Plot of $\ell^2 R/12$ as a function of $y$ and $z$.}
\label{Fig_R1}
\end{figure}

\begin{figure}[tbp]
\centering
\includegraphics[width=8cm]{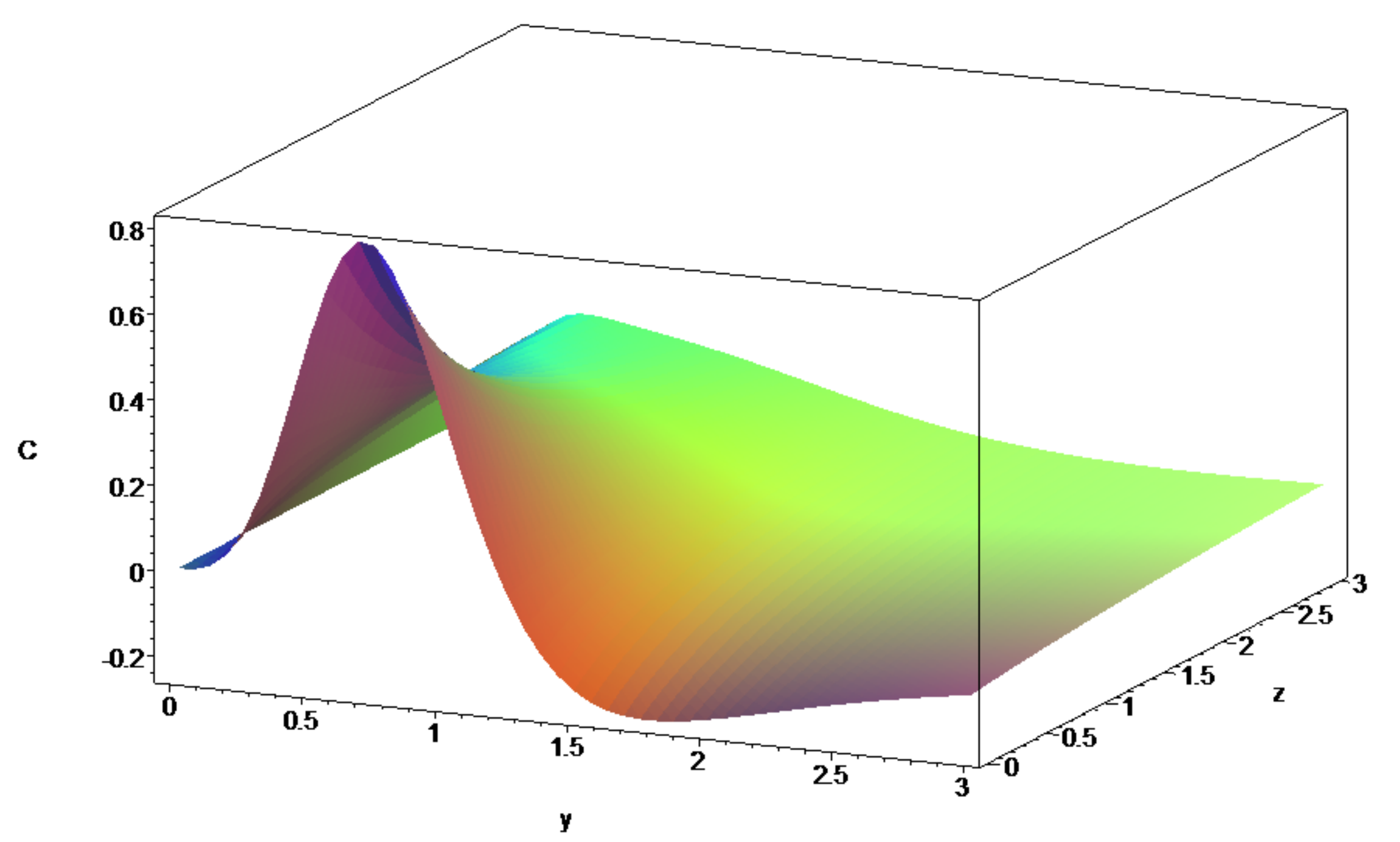}\hfill
  \caption{Plot of $\ell^2 {\cal C}$ as a function of $y$ and $z$.}
\label{Fig_C1}
\end{figure}

\begin{figure}[tbp]
\centering
\includegraphics[width=8cm]{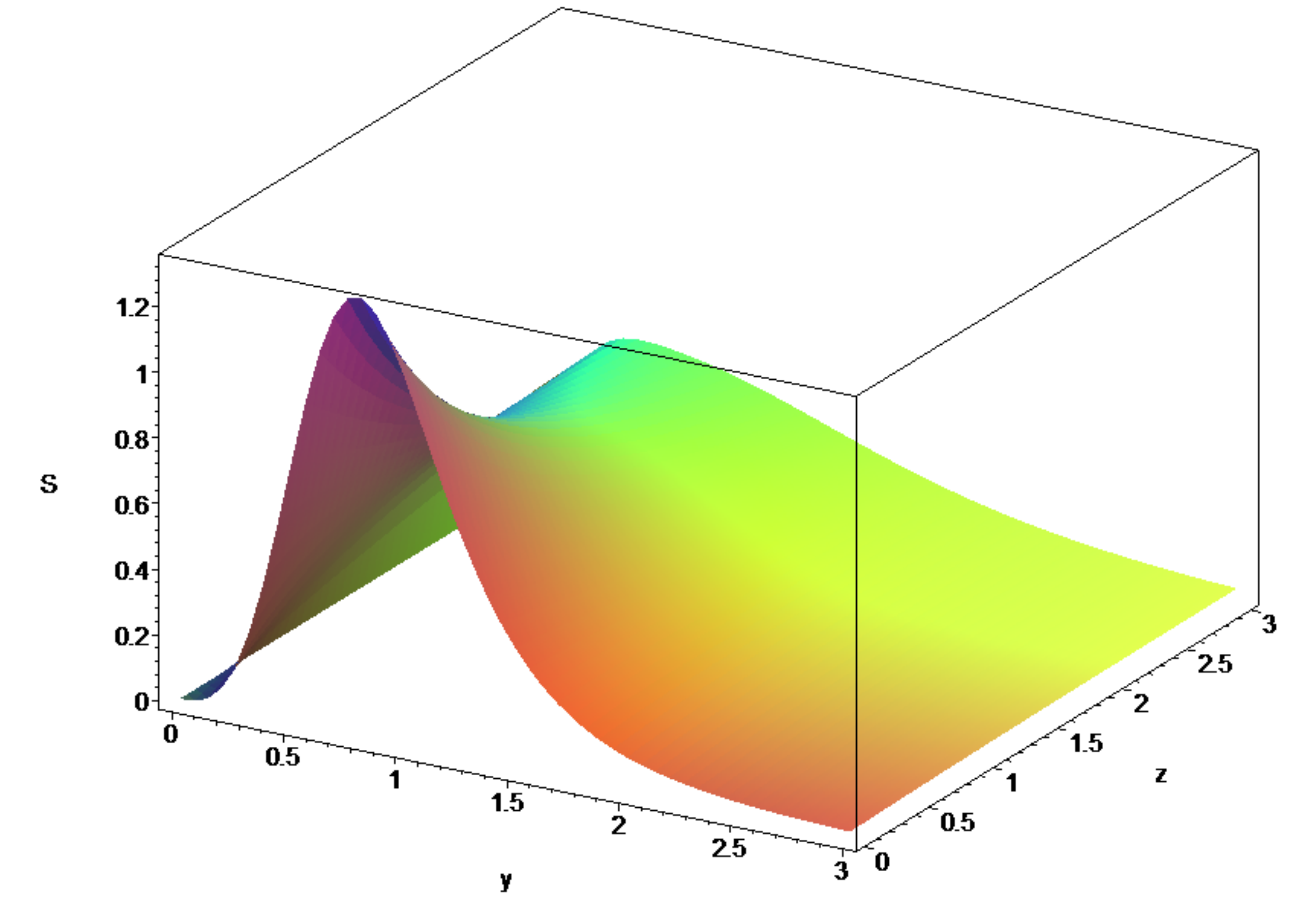}\hfill
  \caption{Plots of  $\ell^2 {\cal S}$  as a function of $y$ and $z$.}
\label{Fig_S1}
\end{figure}

Figures  \ref{Fig_R1}, \ref{Fig_C1} and \ref{Fig_S1}  show plots of the invariants $\ell^2 R/12$, $\ell^2 {\cal C}$ and  $\ell^2 {\cal S}$, respectively. These plots demonstrate that these invariants are uniformly bounded. Hence, the metric \eq{a.7p} satisfies the limiting curvature condition and it can be used as a non-singular model of a black hole.

\subsection{Hayward model}

The metric \eq{a.7p} contains a free dimensionless parameter $c$. It takes a simpler form when this parameter vanishes
\be\n{a.8}
F=1-{2M r^2\over r^3 +2M \ell^2}\, .
\ee
This form of metric for  a non-singular black hole was proposed and discussed in \cite{Hayward:2005gi}.

At large $r$ one has
\BE{a.8a}
F=1-{2M\over r}+O(r^{-4})\, .
\ee

Let us denote by $y$ the dimensionless coordinate
\BE{a.9}
r=(2M\ell^2)^{1/3}y\, .
\ee
Then one has
\ba\n{a.10}
F&=&1-{\cal B}{y^2\over y^3+1}\hhh
R=-{6\over  \ell^2} {y^3-2\over (y^3+1)^3}\, ,\\
{\cal S}&=&{9\over  \ell^2} {y^3\over (y^3+1)^3}\hhh
{\cal C}={\sqrt{12}\over \ell^2} {y^3(y^3-2)\over (y^3+1)^3}\, .
\ea
Here ${\cal B}=(2M/\ell)^{2/3}$.

\begin{figure}[tbp]
\centering
\includegraphics[width=8.5cm]{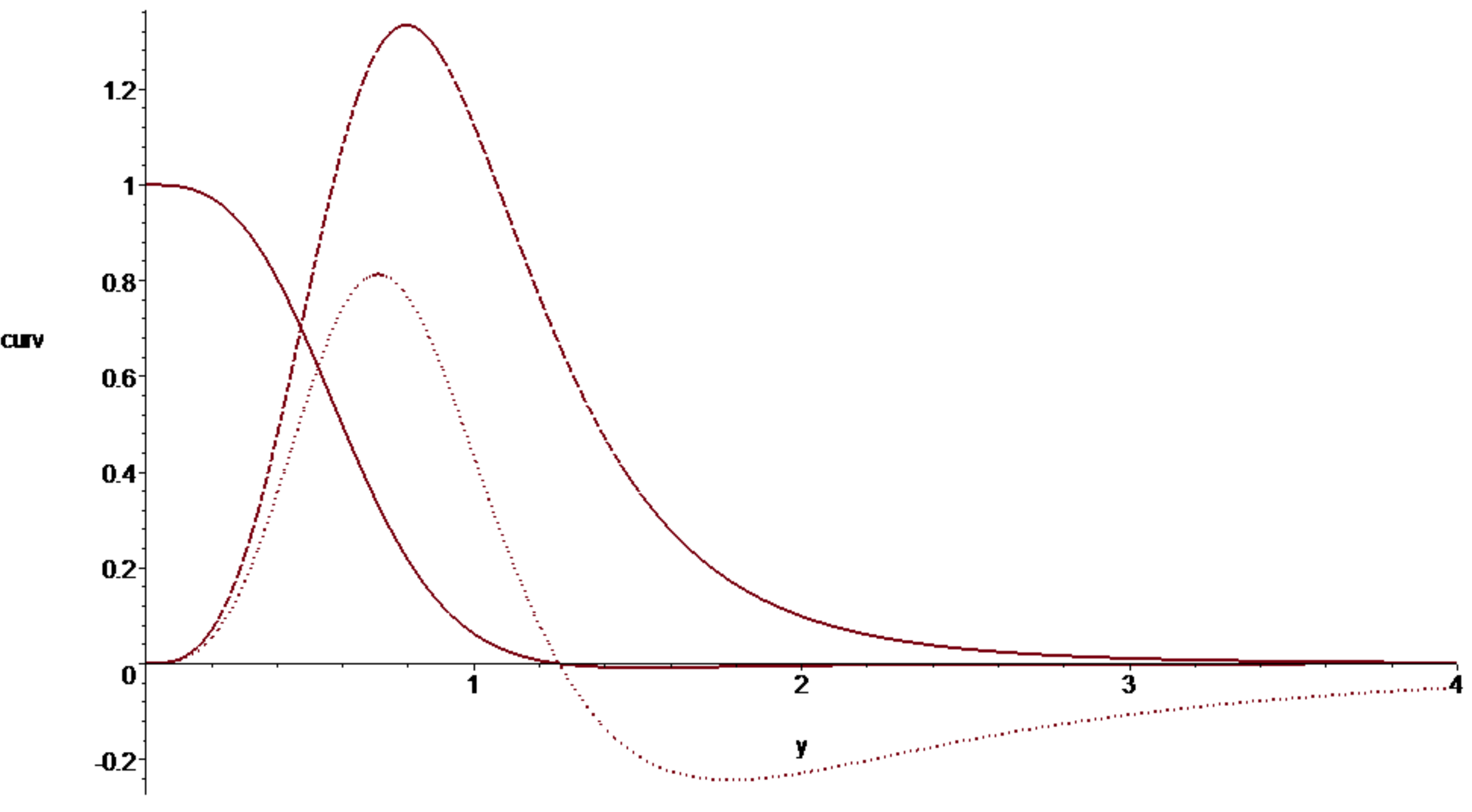}\hfill
  \caption{Plots of $\ell^2 R/12$ (solid line), $\ell^2 {\cal C}$ (dot line) and $\ell^2 {\cal S}$ (dash line) as functions of $y$.}
\label{Fig_4}
\end{figure}

The rational functions of $y$, which enter expressions for $R$, ${\cal S}$ and ${\cal C}$, are regular, finite and their absolute value are restricted by some numerical factor. This means, that that the metric (\ref{a.1}), (\ref{a.8}) satisfies the limiting curvature condition.
The plots of these invariants as functions of $y$ are presented at Figure~\ref{Fig_4}.

The metric (\ref{a.1}), (\ref{a.8}) is invariant under the following scaling transformation
\ba\n{a.11}
&&r\to \alpha r\hhh V\to \alpha V\hhh M\to \alpha M\, ,\nonumber\\
&& \ell\to \alpha \ell\hhh dS^2\to \alpha^2 dS^2\, .
\ea
Using this property one can choose the scale parameter as it is convenient and to work after this with the dimensionless form of the metric. The original metric  (\ref{a.1}), (\ref{a.8}) contains two parameters, the mass $M$ and the fundamental scale $\ell$. However only their (dimensionless) ratio is really important.

The metric  function (\ref{a.8}) can be written in the form
\BE{a.12}
F={(r-r_-)(r-r_+)(r-r_0)\over r^3+B}\, .
\ee
This form of $F$ contains 4 parameters: $r_-$, $r_+$, $r_0$ and $B$. However they are not independent. The condition $F'(0)=0$ gives
\BE{a.13}
r_0=-{r_+ r_-\over r_+ +r_-}\, ,
\ee
while the condition $F(0)=1$ implies
\BE{a.14}
B={r_+^2 r_-^2\over r_+ +r_-}\, .
\ee
Using relations (\ref{a.7}) and (\ref{a.8a}) one finds
\ba\n{a.15}
\ell & = & {r_+ r_-\over \sqrt{r_+^2+r_+ r_- +r_-^2}}\, ,\\
M & = & {r_+^2+r_+ r_- +r_-^2\over 2(r_+ +r_-)} \, .\n{a.15a}
\ea
For given $\ell$ and $M$ one can solve equations \eq{a.15}--\eq{a.15a} and find the radii of the outer and inner horizons, $r_-$ and $r_+$. Solutions exist only if $M\ge M_*=3\sqrt{3} \ell/4$. For this minimal value of mass one has $r_+=r_-=\sqrt{3}\ell$.

There exists another convenient  parametrisation of the metric. One can choose $r_-$ as a scale factor
and define new dimensional coordinates and parameters as follows
\ba\n{a.16}
&&x=r/r_-\, \  p=x_+/r_- \, , m=M/r_- \, ,\nonumber\\
&& b=\ell/r_- \, , v=V/r_- \, .
\ea
One has
\ba
&&dS^2=r_-^2 ds^2\, ,\nonumber\\
&&ds^2=-f dv^2+2 dv dx+x^2 d\omega^2\, ,\n{a.17}\\
&&f={(x-p)(x-1)(x+{p\over p+1})\over x^3+{p^2\over p+1}}\, .\nonumber
\ea
One also has
\BE{a.17S1}
r_-={\ell\over b}\hhh b={p\over \sqrt{p^2+p+1}}\hhh m={p^2+p+1\over 2(p+1)}\, .
\ee

The metric \eq{a.17} has two horizons located at $p$ and $1$. The corresponding dimensionless surface gravity at these horizons is
\be\n{a.17S2}
\kappa_+={(p-1)(p+2)\over 2 p(p^2+p+1)}\, ,\
\kappa_-=-{(p-1)(2p+1)\over 2(p^2+p+1) }\, .
\ee

We denoted by $p$ a position of the outer horizon, so that one has $p\ge 1$.
In the  limit of the large mass, $p\to \infty$, one has
\BE{a.17a}
r_-/\ell\to 1\, ,\  r_+/\ell\to 2m\, ,\  \kappa_+\to {1\over 2p}\, ,\  \kappa_- \to -1\, .
\ee

\subsection{Modified Hayward metric}

In the previous sections we have assumed that the red-shift function $A(r)$ is trivial, $A=1$. This means that there is no frequency shift for the radiation propagating from infinity to the center $r=0$ of the regular black hole. This assumption is rather restrictive. We describe now non-singular black hole models with a non-trivial frequency-shift property. We show that there exist such smooth functions $A(r)$ which produce arbitrary red- or blue-shift effect in the center of the black hole without violation its regularity \footnote{The authors of the paper \cite{DeLorenzo:2014pta} made an attempt to introduce the large frequency shift  by modifying the function $F$ at small radius. They demonstrated that this is impossible without violation of the limiting curvature condition.}.

It is convenient to start with the form \eq{a.17} of the Hayward metric and modify it as follows
\ba
&&dS^2=r_-^2 ds^2\, ,\nonumber\\
&&ds^2=-f A^2 dv^2+2 A dv dx+x^2 d\omega^2\, ,\n{k.1}\\
&&f={(x-p)(x-1)(x+{p\over p+1})\over x^3+{p^2\over p+1}}\, , \nonumber\\
&&A={x^n+1\over x^n+p^k}\, .\nonumber
\ea
Here $n$ and $k$ are properly chosen positive integer numbers. At large distance one has
\BE{k.2}
A\sim 1+{1-p^k\over x^n}+O(x^{-2n})\,
\ee
In order to preserve the correct Schwarzschild asymptotic form one must put $n\ge 2$.
In the presence of the function $A$ the surface gravity \eq{a.17S2} is modified
\ba\n{k.3}
&&\kappa_-\to {2\over p^k+1}\kappa_-\, ,\nonumber\\
&&\kappa_+\to {p^n+1\over  p^n+p^k}\kappa_+=
1-{1-p^k\over 1+p^{n-k}} \kappa_+\, .
\ea
These relations show that for large mass, $p\to\infty$,  the surface gravity at the inner horizon is reduced by the factor $p^{k}$, while $\kappa_+$ remains practically the same.

\begin{figure}[tbp]
\centering
\includegraphics[width=8cm]{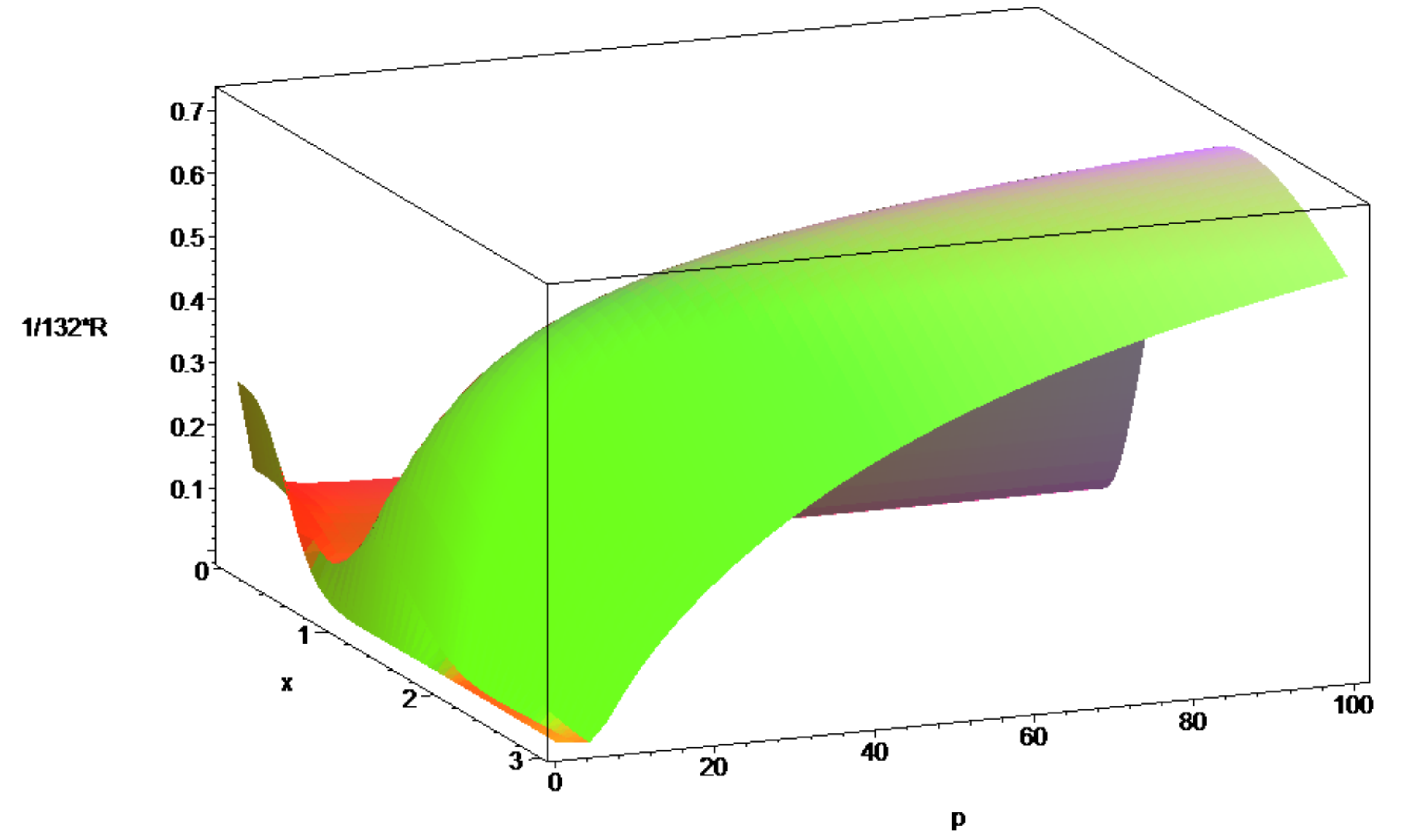}\hfill
\caption{Plot of $r_{-}^2 R/132$ as a function of $x$ and $p$.}
\label{Fig_AR}
\end{figure}

\begin{figure}[tbp]
\centering
\includegraphics[width=8cm]{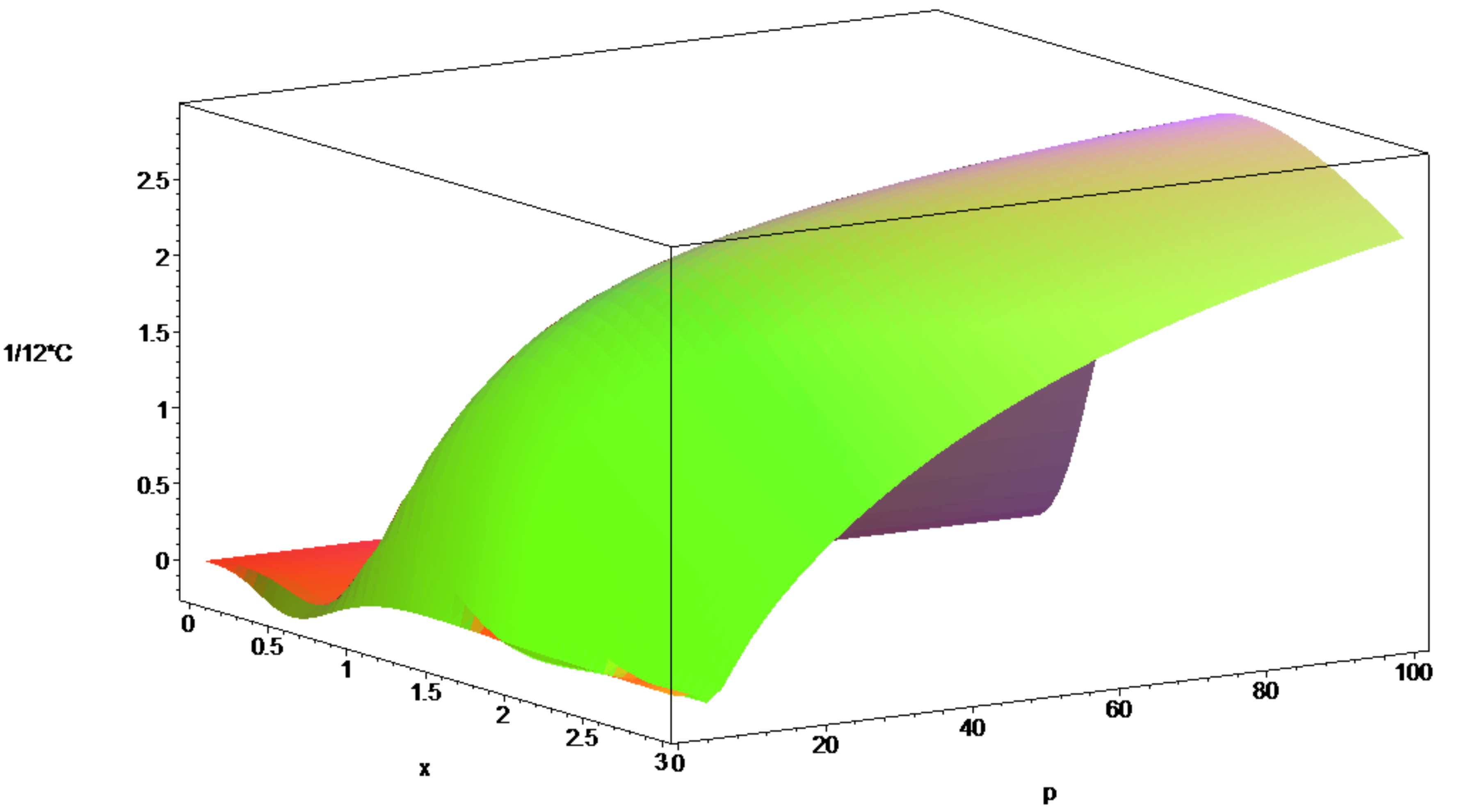}\hfill
  \caption{Plot of $r_{-}^2 {\cal C}/12$ as a function of $x$ and $p$.}
\label{Fig_AC}
\end{figure}

\begin{figure}[tbp]
\centering
\includegraphics[width=8cm]{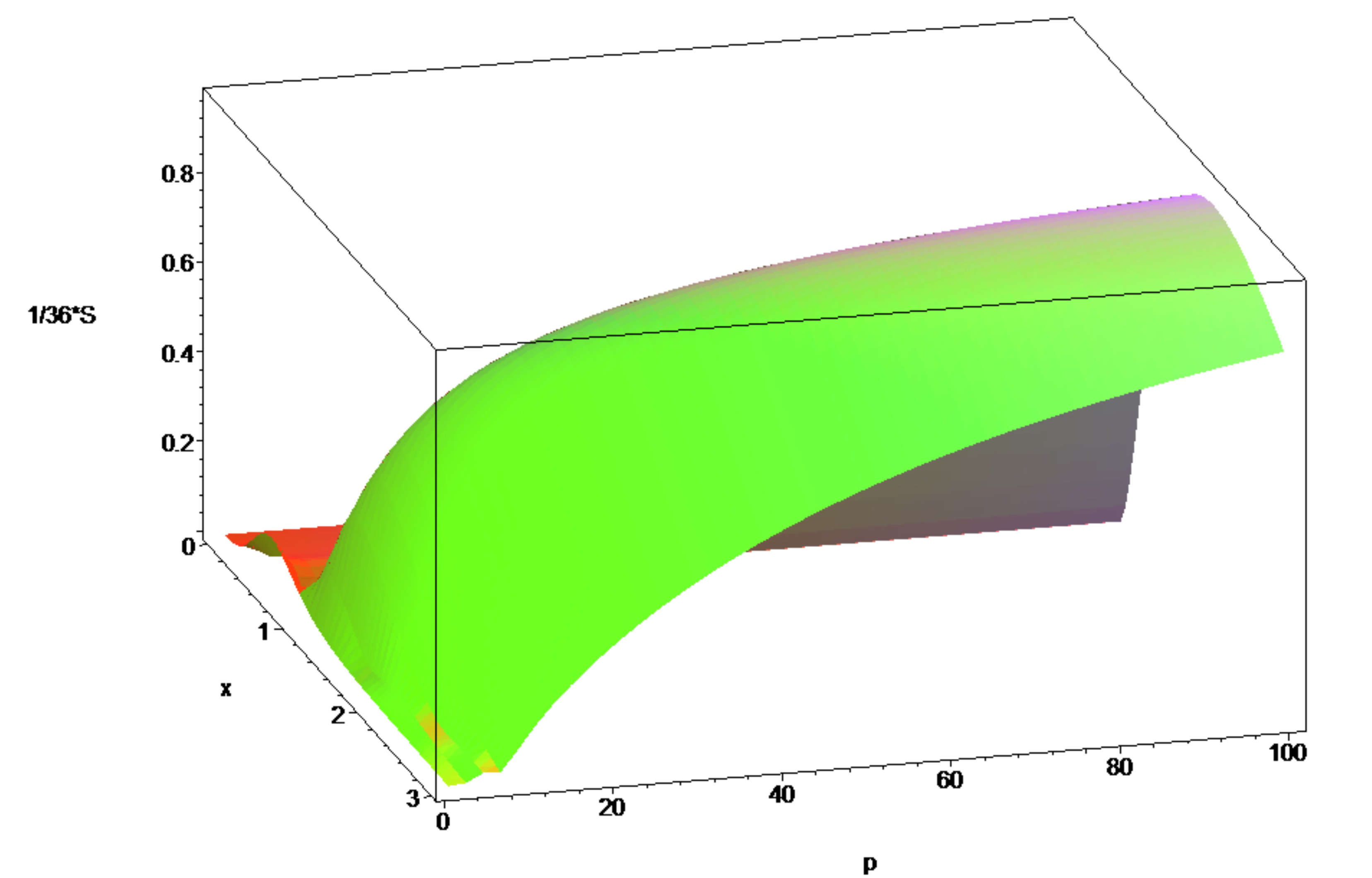}\hfill
  \caption{Plots of  $r_{-}^2 {\cal S}/36$  as a function of $x$ and $p$.}
\label{Fig_AS}
\end{figure}

To illustrate properties of the modified metric we consider a special case $n=6$ and $k=4$
\BE{k.4}
A={x^6+1\over x^6+p^4}\, .
\ee
For this form of the red-shift function, in the limit of large mass ($p\to\infty$),  the surface gravity of the inner horizon becomes small, $\kappa_-\sim p^{-4}$, while the surface gravity $\kappa_+$ remains practically unchanged.

Let us check that the curvature invariants \eq{a.5} for such a metric satisfy the limiting curvature condition. Figures \ref{Fig_AR}--\ref{Fig_AS} show plots $R$, ${\cal C}$ and ${\cal S}$ as functions of $x$ and $p$. They have similar behavior. Namely they are uniformly bounded by numerical factor, independent of $x$ and $p$. In the limit $p\to \infty$ they have the following asymptotic form
\ba\n{k5}
R\sim {12(11 x^6-7 x^4+1)\over x^6+1}+O(p^{-1})\, ,\nonumber\\
{\cal C}\sim {4\sqrt{3} x^4 (7 x^2-4)\over x^6+1}+O(p^{-1})\, ,\\
{\cal S}\sim {6 \sqrt{3} x^4 \sqrt{22x^4-28x^2+9}\over x^6+1}+O(p^{-1})\, .\nonumber
\ea
Hence, the metric \eq{k.1} with non-trivial red-shift property satisfies the limiting curvature condition.

\section{Higher-dimensional non-singular black holes}

The metric \eq{a.1}, \eq{a.8} allows a higher-dimensional generalization. Let us consider static, spherically-symmetric $D=n+1$-dimensional spacetime
\be\n{b.1}
dS^2=-F dV^2+2 dV dr+r^2 d\omega_{n-1}^2\, ,
\ee
where  $F=F(r)$ is of the form
\be\n{b.2}
F=1-{r_g^{n-2} r^2\over r^{n} +r_g^{n-2}\ell^2}\, .
\ee
For $\ell=0$ this metric reproduces the Tangherlini solution of the Einstein equations.
In the four-dimensional case ($n=3$) this metric reduces to \eq{a.1}, \eq{a.8} with $r_g=2M$.
At $r\to \infty$ and $r=0$ one has
\ba\n{b.3}
F&=&1-\left( {r_g\over r}\right)^{n-2} +O(r^{-2(n-1)})\, ,\\
F&=&1-\left( {r\over \ell}\right)^{2} +O(r^{n+2})\, .
\ea

Conditions $F(r*)=F'(r*)=0$ determine the critical value of the gravitational radius
\BE{b.4}
r^*_g=\left( {n\over n-2} \right)^{1/2} \left( {n\over 2}\right)^{1/(n-2)} \ell\, .
\ee
For $r_g> r_g^*$ the metric \eq{b.1}-\eq{b.2} has two horizons, while for $r_g< r_g^*$ the horizons do not exist. For the four-dimensional spacetime, where $n=3$, one reproduces the result of the section $II.2$.

\section{Charged non-singular black hole}

\subsection{Four dimensional spacetime}

Let us consider the metric \eq{a.1} with
\BE{c.1}
F=1-{(2Mr-Q^2)r^2\over r^4 + (2Mr+Q^2)\ell^2}\, .
\ee
In the limit $\ell\to 0$ this metric reproduces the Reissner-Nordrstr\"{o}m metric.
At $r\to \infty$ and $r=0$ one has
\ba\n{c.2}
F&=&1-{2M\over r}+{Q^2\over r^2} +\ell^2 O(r^{-4})\, ,\\
F&=&1+{r^2\over \ell^2} +O(r^6)\, .
\ea
The latter relation means, that the corresponding metric is regular at the origin, where its curvature is of order of $\ell^{-2}$. Deflection of this metric from  the Reissner-Nordrstr\"{o}m at far distance is small and it is controlled by the parameter $\ell^2$.

Let us demonstrate that the metric \eq{c.1} satisfies the limiting curvature property. For this purpose we rewrite $F$ in the form
\BE{c.3}
F=1-{1\over \ell^2} {(ar-z) r^2\over r^4+ar+z}\, ,\ a=2M \ell^2\, ,\ z=Q^2\ell^2\, .
\ee
The calculations give
\ba\n{c.4}
&&R={{\cal A}_R\over \ell^2 N}\hh {\cal S}={{\cal A}_S\over \ell^2 N}\hh
{\cal C}={{\cal A}_C\over \ell^2 N}\, ,\\
&&{\cal A}_R= -6a^2 r^6+12 a^3 r^3+20 r^4 z^2 \nonumber\\
&&\quad \quad +24 a^2 r^2 z+4 a r z^2-12 z^3\, ,\\
&&{\cal A}_S=r(2 r^7 z+9 a^2 r^5+6 a r^4 z \nonumber\\
&&\quad \quad -14 r^3 z^2-2 a^2 r z-4 a z^2)\, ,\n{c.4a}\\
&&{\cal A}_C= 2r(-3 a r^8+6 r^7 z+6 a^2 r^5 \nonumber\\
&&\quad \quad +9 a r^4 z-10 r^3 z^2-2 a z^2)\, ,\\
&& N=(r^4+a r+z)^3 \, .\n{c.4b}
\ea
We use a notation ${\cal A}_{\bullet}$ to denote any of quantities ${\cal A}$ which enter the expressions for the curvature invariants \eq{c.4}--\eq{c.4b}.
Each of ${\cal A}_{\bullet}$ contains three types of the terms: (i) Terms independent of charge; (ii) Terms  independent of mass; and (iii) Terms which depend on both mass and charge. We show now that contribution of each of these types to the curvature invariants obeys an inequality
\BE{c.5}
{|{\cal A}_{\bullet}|\over N}< c_{\bullet}\, ,
\ee
where the constants $c_{\bullet}$ are independent of mass and charge.

For the case (i) one has
\BE{c.6}
{|{\cal A}_{\bullet}(z=0)|\over N}\le {|{\cal A}_{\bullet}(z=0)|\over N(z=0)}\, .
\ee
But the expressions in the right-hand side coincide with similar expressions for the uncharged non-singular black hole, \eq{a.10}, and for this reason they obey the property \eq{c.5}.

For the case (ii) we denote $r=z^{1/4} y$. Then one has
\BE{c.7}
{|{\cal A}_{\bullet}(a=0)|\over N}\le {|{\cal A}_{\bullet}(a=0)|\over N(a=0)}
={|P(y)|\over (y^4+1)^3}\, .
\ee
Here $P(y)$ is a polynomial of $y$ of the power less or equal to 8. Thus the inequality \eq{c.5} is also valid for this contribution.

Let us focus now on the case (iii). Simple analysis shows that there are 3 types of contributions
\BE{c.8}
Q_1={a z^2 r\over N}\hh Q_2={a^2 z r^2\over N}\hh Q_3={a z r^5\over N}\, .
\ee
Since $a\ge 0$ and $z\ge 0$, these functions have similar behavior. They are non-negative and vanish at $r=0$ and $r\to \infty$.
Putting equal to zero the derivatives of these objects with respect to $r$ and solving the obtained relations with respect to $z$ one finds
\BE{c.9}
z_1=r(11 r^3+2 a)\, , \ z_2=r(5 r^3+a/2)\, , \ z_3={r\over 5}(7 r^3-2 a)\, .
\ee
In the last case one should have $7 r^3-2 a>0$. Under these conditions the second derivatives of $Q_i$ with respect to $r$ at the critical points are negative. Thus the functions $Q_i$ have maximum. Putting $r=a^{1/3} u$ one obtain that at the points of their maximum the value of $Q_i$ are
\ba\n{c.10}
\mbox{max}(Q_1)&=&{1\over 27} {(11 u^3+2)^2\over (4 u^3+1)^3}\, ,\\
\mbox{max}(Q_2)&=&{4\over 27} {10 u^3+1\over (4 u^3+1)^3}\, , \\
\mbox{max}(Q_3)&=&{25\over 27} {(7 u^3-2) u^3\over (4 u^3+1)^3}\, .
\ea
This implies that these contributions satisfy \eq{c.5}. Thus we proved that the invariants $|R|$, $|{\cal S}|$ and $|{\cal C}|$ satisfy the limiting curvature condition.

To summarize, the metric \eq{a.1} with the metric function \eq{c.1} describes a non-singular black hole. Its asymptotic at large $r$ correctly reproduces the Reissner-Nordstr\"{o}m metric, so that one can interpret  this metric as a non-singular version of the charged black hole. Certainly, one should assume that besides the gravitational field a system contains also the electromagnetic field, so that the metric \eq{c.1} is a solution of a coupled system of Maxwell and modified gravity equations.

\begin{figure}[tbp]
\centering
\includegraphics[width=8cm]{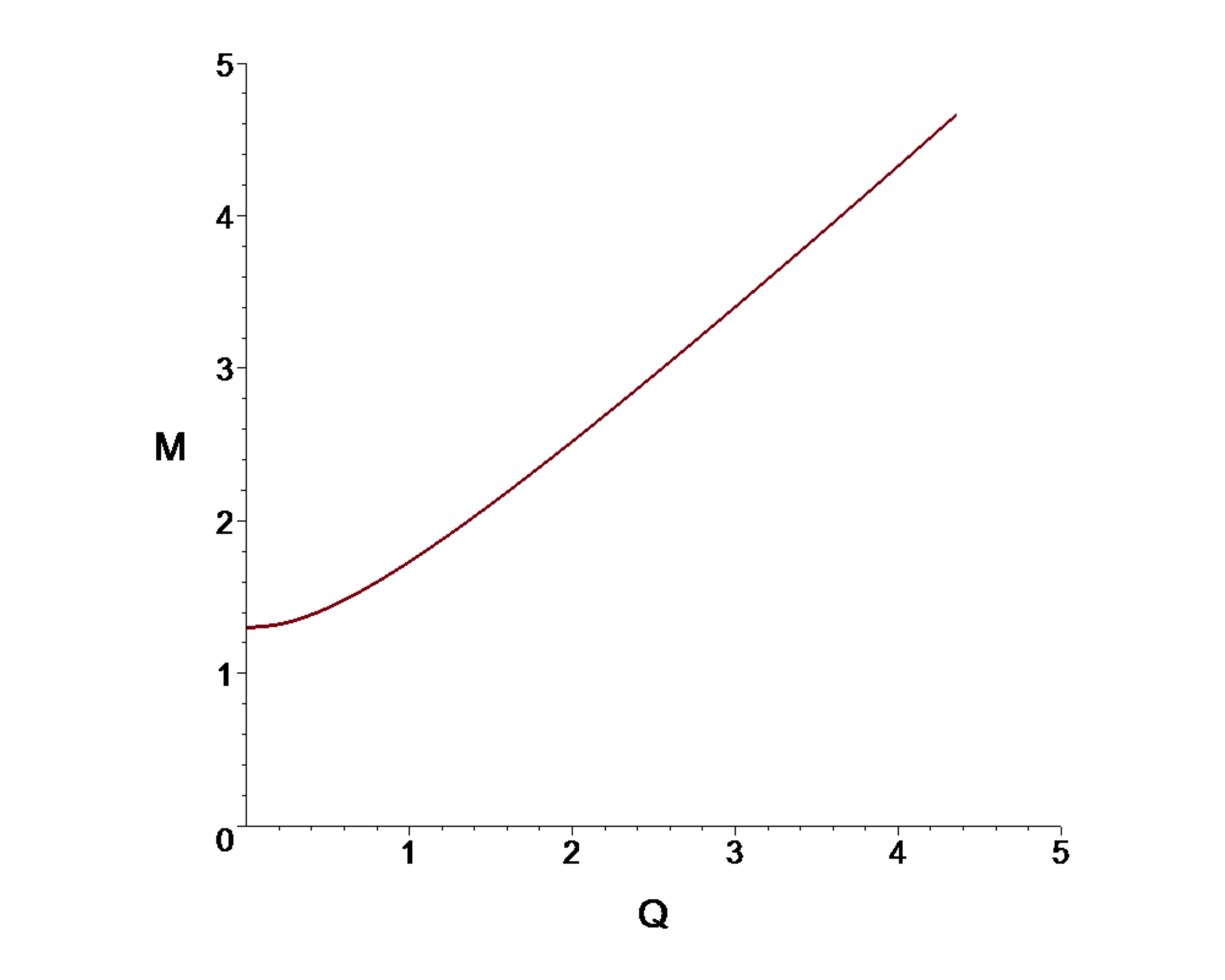}\hfill
  \caption{Critical mass $M/\ell$ for the charged non-singular black hole as a function of $Q/\ell$.}
\label{Fig_MQ}
\end{figure}

Let us find a relation between charge $Q$ and mass $M$ for which the outer and inner horizons coincide. It happens when
\BE{c.11}
F=F'=0\, ,
\ee
where $F$ is given by \eq{c.1} and $F'$ is
\BE{c.12}
F'=-{r(-a r^5+2 a^2 r^2+2 a r z+2 z r^4-2 z^2)\over (r^4+a r+z)^2}\, .
\ee
Solving equation $F=0$ one find
\BE{c.13}
a={(\ell^2 r^4+\ell^2 z+r^2 z)\over r (-\ell^2+r^2)}\, .
\ee
Substituting \eq{c.13} into the equation $F'=0$ and solving it one gets
\BE{c.14}
z={\ell^2 r^4 (r^2-3\ell^2)\over -\ell^4+r^4+4 \ell^2r^2}\, .
\ee
And finally substituting \eq{c.14} one finds
\BE{c.15}
a={2\ell^2(r^2+2\ell^2)r^3\over -\ell^4+r^4+4 \ell^2 r^2}\, .
\ee
Relations \eq{c.14} and \eq{c.15} determined the relation between mass and charge, written in the parametric form. Since both of the quantities $a$ and $z$ should be positive, one has $r\ge \sqrt{3}$.

Plot presented at Figure ~\ref{Fig_MQ} shows critical mass $M$ as a function of charge $Q$.
For small $Q$ one has
\BE{c.20}
M=r_g/2\sim {3\sqrt{3}\over 4} \ell+{1\over \sqrt{3}}{Q^2\over \ell} +O(Q^4)\, .
\ee

\subsection{Higher dimensional generalization}

By comparing \eq{b.2} with \eq{c.1} one can obtain the following higher dimensional version of the charged non-singular black hole. Namely, one uses the form of the metric \eq{b.1} with the metric function of the form
\be\n{c.21}
F=1-{(r_g^{n-2} r^{n-2}-Q^{2(n-2)})r^2\over r^{2(n-2)} +\ell^2 (r_g^{n-1}r^{n-2}+Q^{2(n-2)})}\, .
\ee
This metric in the limit $\ell\to 0$ correctly reproduces the higher dimensional the Reissner-Nordstr\"{o}m metric
\be\n{c.22}
F=1-\left( {r_g\over r}\right)^{n-2}+\left( {Q\over r}\right)^{2(n-2)} +\ell^2 O(r^{2(n-1)})\, .
\ee
It is regular at the origin $r=0$
\BE{c.23}
F=1+{r^2\over \ell^2}+\ldots \, .
\ee

\section{Discussion}

In the present paper we discussed non-singular black hole metrics. We restrict ourselves by metrics which are spherically symmetric and static. Besides "natural" assumptions of regularity at the center $r=0$ and proper asymptotic behavior at the infinity, we required that the corresponding metric also satisfies the limiting curvature condition. The latter condition is rather restrictive and narrows the class of feasible models. The metric proposed by Hayward \cite{Hayward:2005gi} is an important example of the metric for a neutral black hole satisfying these conditions. However this metric has a property which makes it problematic for a self-consistent description of the evaporating black hole. Out-going field modes propagating near the inner horizon are accumulating near it and  experience huge blue shift \cite{Frolov:2014jva}. One can expect that this effect for a quantum field results in the quantum emission of energy \cite{Bolahenko:1986}
\BE{d.1}
\Delta E\sim \exp(-\kappa_- T_{BH})\, .
\ee
Here $\kappa_-$ is the (negative) surface gravity at the inner horizon, which for Hayward model is of the order of	$\ell^{-1}$, and $T_{BH}\sim M^3$ is the life-time of the evaporating black hole. For a self-consistent model of an evaporating black hole one should expect $\Delta E < M$. The expression \eq{d.1} for $M\gg \ell$ does not satisfy this restriction. This indicates that there exists severe self-consistency problem when one  tries to apply the Hayward model to "realistic" quantum black holes.

In the present paper we proposed a class of metrics, which may help to solve this problem. Namely, we used a modification of the metrics with a non-trivial red-shift function $A(r)$ We demonstrate, that this function can be chosen so that the surface gravity $|\kappa_-|$ becomes sufficiently small, so that $\Delta E$, estimated as in \eq{d.1}, can be made rather small.

We also presented a non-singular model for a charged black hole, which obeys the limiting curvature condition. We  briefly discussed higher dimensional versions of such non-singular black holes.
It would be interesting to discuss the application of the presented non-singular metrics for study of the self-consistent models of evaporating black holes. They also might be interesting for a discussion of the information loss paradox. We hope to address these problems in our future work.

%%%%%%%%%%%%%%%%%%%%%%%%%%%%%%%%%%%%%

\section*{Acknowledgments}

The author thanks the Natural Sciences and Engineering Research Council of Canada and the Killam Trust for their financial support.

%%%%%%%%%%%%%%%%%%%%%%%%%%%%%%%%%%%%%%%%%%%%%%%%%%%%%%%%%%%%

%\bibliography{Ghost_references,CHARGE_REG_BH}{}

%merlin.mbs apsrev4-1.bst 2010-07-25 4.21a (PWD, AO, DPC) hacked
%Control: key (0)
%Control: author (0) dotless jnrlst
%Control: editor formatted (1) identically to author
%Control: production of article title (0) allowed
%Control: page (1) range
%Control: year (0) verbatim
%Control: production of eprint (0) enabled
%

\end{document}